\documentclass[aps,twocolumn,showpacs]{revtex4-1}

\usepackage{wallpaper}
\usepackage{dcolumn}
\usepackage{bm}
\usepackage{CJK}
\usepackage{graphics}
\usepackage{multirow}
\usepackage[colorlinks,linkcolor=blue,anchorcolor=blue,citecolor=red]{hyperref}

\begin{document}

\title{Strangeness and $\Delta$ resonance in compact stars with relativistic-mean-field models}
\author{Ting-Ting~Sun$^{1}$}
\author{Shi-Sheng Zhang$^{2}$}
\author{Qiu-Lan~Zhang$^{3}$}
\author{Cheng-Jun~Xia$^{3}$}
\email{cjxia@itp.ac.cn}

\affiliation{$^{1}${School of Physics and Engineering, Zhengzhou University, Zhengzhou 450001, China}
              \\$^{2}${School of Physics and Nuclear Energy Engineering, Beihang University, Beijing 100191, China}
              \\$^{3}${School of Information Science and Engineering, Ningbo Institute of Technology, Zhejiang University, Ningbo 315100, China}}

\date{\today}

\begin{abstract}
We explore the effects of strangeness and $\Delta$ resonance in baryonic matter and compact stars within the relativistic-mean-field
(RMF) models. The covariant density functional PKDD is adopted for $N$-$N$ interaction, parameters fixed based on finite hypernuclei
and neutron stars are taken for the hyperon-meson couplings, and the universal baryon-meson coupling scheme is adopted for the
$\Delta$-meson couplings. In light of the recent observations of GW170817 with the dimensionless combined tidal deformability $197 \leq
\bar{\Lambda}\leq 720$, we find it is essential to include the $\Delta$ resonances in compact stars, and small $\Delta$-$\rho$ coupling
$g_{\rho \Delta}$ is favored if the mass $2.27{}_{-0.15}^{+0.17}\ M_\odot$ of PSR J2215+5135 is confirmed.
\end{abstract}

%21.80.+a Hypernuclei
%26.60.Kp Equations of state of neutron-star matter
%97.60.Jd Neutron star core

\maketitle

\section{\label{sec:intro}Introduction}
The recent observation of gravitational waves from the binary neutron star merger event GW170817 suggests that the merging objects
are compact~\cite{LIGO2017_PRL119-161101, LigoVirgo2018_arXiv1805.11581}. Assuming low spin priors, the dimensionless combined tidal
deformability $\bar{\Lambda}$ is considered to be less than 720 at 90\% confidence level~\cite{LigoVirgo2018_arXiv1805.11579},
while a lower limit with $\bar{\Lambda}\geq 197$ is obtained based on electromagnetic observations of the transient counterpart
AT2017gfo~\cite{Coughlin2018_arXiv1805.09371}. Even though the observations of neutron stars' radii are controversial and
depend on specific assumptions, the recent measurements seem to be converging and lie at the lower end of 10-14 km
range~\cite{Guillot2013_ApJ772-7, Lattimer2014_EPJA50-40, Ozel2016_ARAA54-401, Li2015_ApJ798-56, Steiner2018_MNRAS476-421, Most2018_PRL120-261103,
LigoVirgo2018_arXiv1805.11581}. The combined constraints on the tidal deformability and radii of neutron stars indicate a
soft equation of states (EoS), where many covariant density functionals are in jeopardy~\cite{Zhu2018_ApJ862-98,
Malik2018_PRC98-035804}. A possible solution to this problem is to introduce new degrees of freedom, e.g., $\Delta$
resonances, hyperons, and deconfined quarks~\cite{Gomes2018_arXiv1806.04763}. As one increases the density of nuclear matter,
the inevitable emergence of $\Delta$ isobars, hyperons, and quarks can soften the EoSs significantly and reduce the radius and
tidal deformability of the corresponding compact stars, which can be consistent with these recent observations.

However, a soft EoS will result in compact stars with too small masses that can not reach two solar mass as observed in pulsars
PSR J1614-2230 ($1.928 \pm 0.017\ M_\odot$)~\cite{Demorest2010_Nature467-1081, Fonseca2016_ApJ832-167} and PSR J0348+0432
($2.01 \pm 0.04\ M_\odot$)~\cite{Antoniadis2013_Science340-1233232}, i.e., the Hyperon Puzzle~\cite{Vidana2015_AIPCP1645-79}
or $\Delta$ Puzzle~\cite{Drago2014_PRC90-065809}. Extensive efforts were
made to resolve the Hyperon Puzzle~\cite{Weissenborn2012_PRC85-065802, Bednarek2012_AA543-A157, Oertel2015_JPG42-075202,
Maslov2015_PLB748-369, Maslov2016_NPA950-64, Takatsuka_EPJA13-213, Vidana2011_EPL94-11002, Yamamoto2013_PRC88-022801,
Lonardoni2015_PRL114-092301, Togashi2016_PRC93-035808, Weissenborn2011_ApJ740-L14, Klahn2013_PRD88-085001, Zhao2015_PRD92-054012,
Kojo2015_PRD91-045003, Masuda2016_EPJA52-65, Li2015_PRC91-035803, Whittenbury2016_PRC93-035807, Fukushima2016_ApJ817-180,
Sun2018_CPC42-25101} and $\Delta$ Puzzle~\cite{Drago2014_PRD89-043014, Cai2015_PRC92-015802, Zhu2016_PRC94-045803,
Bai2018_PRD97-023018}. Nevertheless, with the constrained observable tidal deformability of GW170817~\cite{LIGO2017_PRL119-161101,
LigoVirgo2018_arXiv1805.11579, Coughlin2018_arXiv1805.09371}, those solutions may be challenged, especially for the latest observation
of a more massive PSR J2215+5135 ($2.27{}_{-0.15}^{+0.17}\ M_\odot$)~\cite{Linares2018_ApJ859-54}.

To satisfy these stringent observational constraints, we consider the possible existence of both $\Delta$ isobars
and hyperons in neutron stars. Since relativistic-mean-field (RMF) models~\cite{Brockmann1977_PLB69-167, Boguta1981_PLB102-93,
Mares1989_ZPA333-209, Mares1994_PRC49-2472, Toki1994_PTP92-803, Song2010_IJMPE19-2538, Tanimura2012_PRC85-014306, Wang2013_CTP60-479}
have been successfully adopted to describe finite (hyper)nuclei~\cite{Reinhard1989_RPP52-439, Ring1996_PPNP37_193-263,
Meng2006_PPNP57-470, Paar2007_RPP70-691, Meng2015_JPG42-093101, Meng2016_RDFNS, Typel1999_NPA656-331, Vretenar1998_PRC57-R1060,
Lu2011_PRC84-014328, Hagino2014_arXiv1410.7531, Sun2016_PRC94-064319} and baryonic matter~\cite{Glendenning2000, Ban2004_PRC69-045805,
Weber2007_PPNP59-94, Long2012_PRC85-025806, Sun2012_PRC86-014305, Wang2014_PRC90-055801, Fedoseew2015_PRC91-034307, Gao2017_ApJ849-19},
in this work the EoSs of baryonic matter are obtained based on RMF model. More specifically, we adopt the covariant density
functional PKDD~\cite{Long2004_PRC69-034319}, while the hyperon-meson couplings are fixed based on our
previous investigations on hypernuclei and neutron stars~\cite{Sun2016_PRC94-064319, Sun2018_CPC42-25101, Liu2018_PRC98-024316}.
For the $\Delta$-meson couplings, as in Ref.~\cite{Drago2014_PRC90-065809}, we adopt the universal baryon-meson
coupling scheme, while a vanishing $\Delta$-$\rho$ coupling is considered as well.
It is found that the observational tidal deformability and mass of PSR J2215+5135 can be reproduced only by including $\Delta$
isobars in neutron stars.

The paper is organized as follows. In Sec.~\ref{sec:the}, we present the formalism of RMF model for baryonic matter,
the choices of baryon-meson couplings, the conditions for obtaining the EoSs of neutron star matter,
and the formalism to determine the structures of compact stars. Results and discussions are given in Sec.~\ref{sec:num}.
We make a summary in Sec.~\ref{sec:con}.

\section{\label{sec:the}Theoretical framework}
The Lagrangian density of RMF models is given as
\begin{eqnarray}
\mathcal{L}
 &=& \sum_{\mathrm{b}} \bar{\psi}_\mathrm{b}
       \left[  i \gamma^\mu \partial_\mu- m_\mathrm{b} - g_{\sigma\mathrm{b}}\sigma
              - g_{\omega\mathrm{b}} \gamma^\mu \omega_\mu
         \right.
\nonumber \\
 &&\mbox{}\left.- g_{\rho\mathrm{b}} \gamma^\mu \boldsymbol{\tau}_\mathrm{b} \cdot \boldsymbol{\rho}_\mu
              - \gamma^\mu A_\mu q_\mathrm{b}
              \right] \psi_\mathrm{b} + \frac{1}{2}\partial_\mu \sigma \partial^\mu \sigma
\nonumber \\
 &&\mbox{}   - \frac{1}{2}m_\sigma^2 \sigma^2 - \frac{1}{4} \omega_{\mu\nu}\omega^{\mu\nu} + \frac{1}{2}m_\omega^2 \omega_\mu\omega^\mu
\nonumber \\
 &&\mbox{} - \frac{1}{4} \boldsymbol{\rho}_{\mu\nu}\cdot\boldsymbol{\rho}^{\mu\nu}
     + \frac{1}{2}m_\rho^2 \boldsymbol{\rho}_\mu\cdot\boldsymbol{\rho}^\mu - \frac{1}{4} A_{\mu\nu}A^{\mu\nu}
\nonumber \\
 &&\mbox{}
     +\sum_{l=e,\mu} \bar{\psi}_l \left[ i \gamma^\mu \partial_\mu - m_l + e \gamma^\mu A_\mu \right]\psi_l,
\label{eq:Lagrange}
\end{eqnarray}
with the field tensors
\begin{eqnarray}
\omega_{\mu\nu} &=& \partial_\mu \omega_\nu - \partial_\nu \omega_\mu, \nonumber\\
\boldsymbol{\rho}_{\mu\nu}
  &=& \partial_\mu \boldsymbol{\rho}_\nu - \partial_\nu \boldsymbol{\rho}_\mu, \\
A_{\mu\nu} &=& \partial_\mu A_\nu - \partial_\nu A_\mu.\nonumber
\end{eqnarray}
The included baryons here are nucleons, hyperons ($\Lambda^0$, $\Sigma^{+,0,-}$, and $\Xi^{0,-}$), and $\Delta$ resonance.
To describe the baryon-baryon interactions, the isoscalar-scalar channel ($\sigma$), isoscalar-vector channel ($\omega$)
and isovector-vector channel ($\boldsymbol{\rho}$) are considered.

Based on the Typel-Wolter ansatz~\cite{Typel1999_NPA656-331}, the density dependence of coupling constants
$g_{\xi\mathrm{b}}~(\xi=\sigma$, $\omega$)
are obtained with
\begin{equation}
g_{\xi\mathrm{b}}(n) = g_{\xi\mathrm{b}}(n_0) a_{\xi} \frac{1+b_{\xi}(n/n_0+d_{\xi})^2}
                          {1+c_{\xi}(n/n_0+e_{\xi})^2}, \label{eq:ddcp_TW}
\end{equation}
where $n$ is the density of nuclear matter with $n_0$ being the saturation density. Note that a different formula is adopted
for the $\boldsymbol{\rho}$ meson, i.e.,
\begin{equation}
g_{\rho\mathrm{b}}(n) = g_{\rho\mathrm{b}}(n_0) \exp{\left[-a_\rho(n/n_0-1)\right]}. \label{eq:ddcp_rho}
\end{equation}

For a system with time-reversal symmetry, the space-like components of the vector fields $\omega_\mu$ and $\boldsymbol{\rho}_\mu$ vanish,
leaving only the time components $\omega_0$ and $\boldsymbol{\rho}_0$. Meanwhile, the charge conservation guarantees that only the 3rd
component in the isospin space of $\boldsymbol{\rho}_0$ survives. In the mean-field and no-sea approximations, the single particle
(s.p.) Dirac equations for baryons and the Klein-Gordon equations for mesons and photon can be obtained from the variational procedure.

For the $N$-$N$ interactions, we adopt the covariant density functional PKDD~\cite{Long2004_PRC69-034319}, which gives the
saturation density $n_0=0.149552\ \mathrm{fm}^{-3}$, saturation energy $E_0=-16.267$ MeV, incompressibility $K=262.181$ MeV and
symmetry energy $E_\mathrm{sym}=36.790$ MeV.

Beside nucleons, we also consider the effects of strangeness and $\Delta$ resonance, i.e., $\Lambda$, $\Xi$, $\Sigma$, and
$\Delta$ baryons. For the $\Lambda$-$\omega$ coupling, according to our previous investigations~\cite{Sun2018_CPC42-25101},
the mass of PSR J0348+0432 can only be attained with large values of $g_{\omega \Lambda}$ at fixed $\Lambda$ potential
well depth ($V_\Lambda = - 29.786$~MeV) in symmetric nuclear matter ($n_p=n_n = n_0/2$). Thus, in this work we suppose $g_{\omega \Lambda}
= g_{\omega N}$, which gives $g_{\sigma \Lambda} = 0.878 g_{\sigma N}$. Similarly, we fix the $\Xi$-meson and $\Sigma$-meson
couplings with $g_{\omega \Xi} = g_{\omega \Sigma} = g_{\omega N}$, $g_{\sigma \Xi} =0.844 g_{\sigma N}$, and $g_{\sigma \Sigma} =
0.878 g_{\sigma N}$, which corresponds to the potential well depths $V_\Xi = - 16.276$~MeV and $V_\Sigma = - 29.957$
MeV~\cite{Liu2018_PRC98-024316}. Note that there are some ambiguity on the potential well depth $V_\Sigma$, where the $(\pi^-, K^+)$
reactions on medium-to-heavy nuclei indicates a repulsive potential~\cite{Noumi2002_PRL89-072301, Noumi2003_PRL90-049902,
Saha2004_PRC70-044613, Kohno2006_PRC74-064613} while the observation of a ${}_\Sigma^4\mathrm{He}$ bound state in the
$(K^-, \pi^-)$ reaction favors an attractive potential~\cite{Nagae1998_PRL80-1605}. For the hyperon-$\rho$ couplings, we take
$g_{\rho \Lambda} = 0$ and $g_{\rho \Xi} = g_{\rho \Sigma} = g_{\rho N}$ according to their isospin characters~\cite{Sun2016_PRC94-064319,
Liu2018_PRC98-024316}. In principle, in consideration of the hyperon-hyperon interactions such as the weakly attractive $\Lambda$-$\Lambda$
interaction, the exchange of $\sigma^*$ and $\phi$ mesons between hyperons should also be taken into account. However, according
to the recipe of various baryon-meson couplings inspired by the symmetries of the baryon octet~\cite{Schaffner1996_PRC53-1416,
Weissenborn2012_PRC85-065802, Miyatsu2013_PRC88-015802, Oertel2015_JPG42-075202}, taking $g_{\omega \Lambda} = g_{\omega \Xi} =
g_{\omega \Sigma} = g_{\omega N}$ and $g_{\phi N} = 0$ indicates vanishing hyperon-$\phi$ couplings. In such cases, the contributions
from $\sigma^*$ and $\phi$ mesons are neglected in our Lagrange density~(\ref{eq:Lagrange}).

For the $\Delta$-$\omega$ and $\Delta$-$\sigma$ couplings, they are often chosen to be close to the $N$-$\omega$ and $N$-$\sigma$
couplings, i.e., $g_{\omega \Delta}\approx g_{\omega N}$ and $g_{\sigma \Delta} \approx g_{\sigma N}$~\cite{Li1997_PRC56-1570,
Kosov1998_PLB421-37, Drago2014_PRC90-065809, Drago2014_PRD89-043014, Zhu2016_PRC94-045803}, which can be attributed
to the similar potential depths of $\Delta$s and nucleons in nuclear medium according to the data analyses of photoabsorption,
electron-nucleus, and pion-nucleus scattering~\cite{Drago2014_PRC90-065809}. Slight deviations from those
values were also explored in Ref.~\cite{Schdoturhoff2010_ApJ724-L74}. However, little is known for the $\Delta$-$\rho$ coupling,
while the linear dependence of the onset density $n_{\Delta^-}^\mathrm{crit}$ with $g_{\rho \Delta}$ was reported in
Refs.~\cite{Cai2015_PRC92-015802, Zhu2016_PRC94-045803}. Therefore, in this work we adopt the universal baryon-meson coupling
scheme with $g_{\omega \Delta}= g_{\omega N}$, $g_{\sigma \Delta} = g_{\sigma N}$, and $g_{\rho \Delta}= g_{\rho N}$. To see the
possibility of smaller $g_{\rho \Delta}$, we also study the cases with $g_{\rho \Delta}=0$. Since the $\Delta$ baryons
have a Breit-Wigner mass distribution around the centroid mass 1232 MeV with a width of about 120 MeV, the variation of $m_\Delta$
has sizable effects on baryonic matter and structures of compact stars~\cite{Cai2015_PRC92-015802}. In this work, we adopt
various $\Delta$ masses with $m_\Delta = 1112$~MeV, $1232$~MeV, and $1352$~MeV.

\begin{table}[h]
\centering
\caption{\label{Tab:by_list}Strangeness number $S$, mass $M$, third component of isospin $\tau_{3}$, total angular momentum and parity $J^{P}$,
charge $q$, and coupling constants $\alpha_\xi=g_{\xi\mathrm{b}}/g_{\xi N}$ ($\xi=\sigma, \omega$, and $\rho$) for $\Lambda^0$, $\Xi^{0,-}$,
and $\Sigma^{+,0,-}$ hyperons and $\Delta$ baryons.}
\begin{tabular}{l|ccccc|ccc}
\hline\hline
              & $S$  & $M$ (MeV)     & $\tau_{3}$ & $J^{p}$     & $q$ ($e$) & $\alpha_\sigma$ & $\alpha_\omega$ & $\alpha_\rho$    \\ \hline
 $\Lambda^0$  & $-1$ & $1115.6$      &  $0$       & $(1/2)^{+}$ & $0$     &  0.878  & 1 &  0    \\ \hline
 $\Xi^0$      & $-2$ & $1314.9$      &  $+1$      & $(1/2)^{+}$ & $0$     &  0.844  & 1 &  1    \\
 $\Xi^-$      & $-2$ & $1321.3$      &  $-1$      & $(1/2)^{+}$ & $-1$    &  0.844  & 1 &  1    \\  \hline
 $\Sigma^+$   & $-1$ & $1189.4$      &  $+1$      & $(1/2)^{+}$ & $+1$    &  0.878  & 1 &  1    \\
 $\Sigma^0$   & $-1$ & $1192.5$      &  $0$       & $(1/2)^{+}$ & $0$     &  0.878  & 1 &  1    \\
 $\Sigma^-$   & $-1$ & $1197.4$      &  $-1$      & $(1/2)^{+}$ & $-1$    &  0.878  & 1 &  1    \\ \hline
$\Delta^{++}$ & $0$  & $1232\pm 120$ &  $+3$      & $(3/2)^{+}$ & $+2$    &     1   & 1 & 0, 1  \\
 $\Delta^{+}$ & $0$  & $1232\pm 120$ &  $+1$      & $(3/2)^{+}$ & $+1$    &     1   & 1 & 0, 1  \\
 $\Delta^0$   & $0$  & $1232\pm 120$ &  $0$       & $(3/2)^{+}$ & $0$     &     1   & 1 & 0, 1  \\
 $\Delta^-$   & $0$  & $1232\pm 120$ &  $-3$      & $(3/2)^{+}$ & $-1$    &     1   & 1 & 0, 1  \\
  \hline\hline
\end{tabular}
\end{table}

In Table~\ref{Tab:by_list}, we list properties and coupling constants for baryons other than nucleons in Eq.~(\ref{eq:Lagrange}).
Meanwhile, it is worth mentioning that the covariant density functional PKDD adopted here is phenomenological, where the
nucleon-meson coupling constants are fixed according to the masses of spherical nuclei, the incompressibility, saturation density,
and symmetry energy of nuclear matter~\cite{Long2004_PRC69-034319}. In light of the recent developments of microscopic many-body
calculations in describing finite nuclei and nuclear matter starting from realistic nucleon-nucleon
interactions~\cite{Lynn2016_PRL116-062501, Hagen2016_PRL117-172501, Hergert2017, Simonis2017_PRC96-014303, Holt2017_PRC95-034326,
Meisner2016_PS91-033005, Hu2017_PRC96-034307}, a more refined adjustment of parameters incorporating those results are necessary.
A possible way to reach this in RMF model is to introduce density-dependent coupling constants derived from self-energies of
Dirac-Brueckner calculations of nuclear matter~\cite{Typel1999_NPA656-331, Roca-Maza2011_PRC84-054309}, which are found decreasing
with density and can be reproduced with Eqs.~(\ref{eq:ddcp_TW}) and (\ref{eq:ddcp_rho}).

Based on the Lagrangian density in Eq.~(\ref{eq:Lagrange}), the meson fields are obtained by solving
\begin{eqnarray}
m_\sigma^2 \sigma &=& -\sum_\mathrm{b} g_{\sigma\mathrm{b}} n_\mathrm{b}^\mathrm{s}, \label{eq:eom_sigma} \nonumber\\
m_\omega^2 \omega_0 &=& \sum_\mathrm{b} g_{\omega\mathrm{b}} n_\mathrm{b},  \label{eq:eom_omega}\\
m_\rho^2 \rho_{0,3} &=& \sum_\mathrm{b} g_{\rho\mathrm{b}}\tau_{\mathrm{b},3} n_\mathrm{b}, \nonumber \label{eq:eom_rho}
\end{eqnarray}
with the number density $n_\mathrm{b}=\langle \bar{\psi}_\mathrm{b}\gamma^0 \psi_\mathrm{b} \rangle$ and scalar density
$n_\mathrm{b}^\mathrm{s}=\langle \bar{\psi}_\mathrm{b}\psi_\mathrm{b} \rangle$ of baryon type b, which are given in
Eqs.~(\ref{eq:ni}) and (\ref{eq:ns}). Here we take $\sigma$, $\omega_0$
and $\rho_{0,3}$ as their mean values.

At zero temperature, with no sea approximation, the energy density can be determined by
\begin{eqnarray}
E &=& \sum_{i={\mathrm{b},l}}\varepsilon_i(\nu_i, m_i^*) + \sum_{\xi=\sigma, \omega, \boldsymbol{\rho}} \frac{1}{2}m_\xi^2 \xi^2,
\label{eq:E}
\end{eqnarray}
in which the kinetic energy density of fermion $i$ is
\begin{eqnarray}
\varepsilon_i(\nu_i, m_i) &=& \int_0^{\nu_i} \frac{f_i p^2}{2\pi^2} \sqrt{p^2+m_i^2}\mbox{d}p \label{eq:ei0}\\
&=&  \frac{f_i m_i^4}{16\pi^{2}} \left[x_i(2x_i^2+1)\sqrt{x_i^2+1}-\mathrm{arcsh}(x_i) \right].
\nonumber
\end{eqnarray}
Here we have defined $x_i\equiv \nu_i/m_i$ with $\nu_i$ being the Fermi momentum and $f_i = 2 J_i +1$ the degeneracy factor of
particle type $i$. Note that in Eq.~(\ref{eq:E}), the baryon effective mass is defined as $m_\mathrm{b}^*\equiv m_\mathrm{b} +
g_{\sigma\mathrm{b}} \sigma$, while the mass of leptons remain constants with $m_l^*\equiv m_l$.
The source currents of fermion $i$ are given by
\begin{eqnarray}
n_i &=& \langle \bar{\psi}_i\gamma^0 \psi_i \rangle = \frac{f_i\nu_i^3}{6\pi^2}, \label{eq:ni} \\
n_i^\mathrm{s}&=&\langle \bar{\psi}_i\psi_i \rangle
=\frac{f_i m_i^3}{4\pi^2} \left[x_i\sqrt{x_i^2+1} - \mathrm{arcsh}(x_i)\right]. \label{eq:ns}
\end{eqnarray}
The chemical potentials for baryons $\mu_\mathrm{b}$ and leptons $\mu_l$ are
\begin{eqnarray}
\mu_\mathrm{b}&=& g_{\omega\mathrm{b}} \omega_0
              + g_{\rho\mathrm{b}}\tau_{\mathrm{b},3} \rho_{0,3}
              + \Sigma^\mathrm{R}_{\mathrm{b}}
              + \sqrt{\nu_\mathrm{b}^2+{m_\mathrm{b}^*}^2},
\label{eq:chem_B} \\
\mu_l &=&  \sqrt{\nu_l^2+m_l^2},
\label{eq:chem_l}
\end{eqnarray}
with the ``rearrangement" term
\begin{eqnarray}
&&\Sigma^\mathrm{R}_{\mathrm{b}}=\nonumber\\
&& \sum_{\mathrm{b}}\left(
   \frac{\mbox{d} g_{\sigma \mathrm{b}}}{\mbox{d} n} \sigma n_\mathrm{b}^\mathrm{s}+
   \frac{\mbox{d} g_{\omega\mathrm{b}}}{\mbox{d} n} \omega_0 n_\mathrm{b}+
   \frac{\mbox{d} g_{\rho\mathrm{b}}}{\mbox{d} n} \rho_{0,3}\tau_{\mathrm{b},3} n_\mathrm{b}
  \right).
\label{eq:re_B}
\end{eqnarray}
Then the pressure is expressed by
\begin{equation}
P = \sum_i \mu_i n_i \nonumber - E. \label{eq:pressure}
\end{equation}

For neutron star matter, it should fulfill the charge neutrality condition
\begin{equation}
  \sum_i q_i n_i = 0, \label{eq:Chntr}
\end{equation}
with $q_i$ being the charge of particle type $i$. To reach the lowest energy, particles will undergo weak reactions until
the $\beta$-equilibrium condition is satisfied, i.e.,
\begin{equation}
\mu_\mathrm{b}= \mu_n - q_\mathrm{b} \mu_e,~~ \mu_\mu = \mu_e.  \label{eq:weakequi}
\end{equation}
The EoS of neutron star matter can be obtained from Eqs.~(\ref{eq:E}) and (\ref{eq:pressure}), which is the input of the
Tolman-Oppenheimer-Volkov (TOV) equation
\begin{equation}
\frac{\mbox{d}P}{\mbox{d}r}
=-\frac{G M E}{r^2}
  \frac{(1+P/E)(1+4\pi r^3 P/M)} {1-2G M/r}.  \label{eq:TOV}
\end{equation}
By solving the TOV equation with the subsidiary condition
\begin{equation}
\frac{\mbox{d}M(r)}{\mbox{d}r} = 4\pi E r^2, \label{eq:m_star}
\end{equation}
we get the relation of mass $M$ and radius $R$ of a neutron star.
Here, the gravity constant $G=6.707\times 10^{-45}\ \mathrm{MeV}^{-2}$.
The tidal deformability of a compact star is extracted from
\begin{equation}
\Lambda = \frac{2 k_2}{3}\left( \frac{R}{G M} \right)^5, \label{eq:td}
\end{equation}
where $k_2$ is the second Love number and can be fixed simultaneously with the structures of compact stars~\cite{Damour2009_PRD80-084035,
Hinderer2010_PRD81-123016, Postnikov2010_PRD82-024016}.

\section{\label{sec:num}Results and discussions}

\begin{figure}
\includegraphics[width=\linewidth]{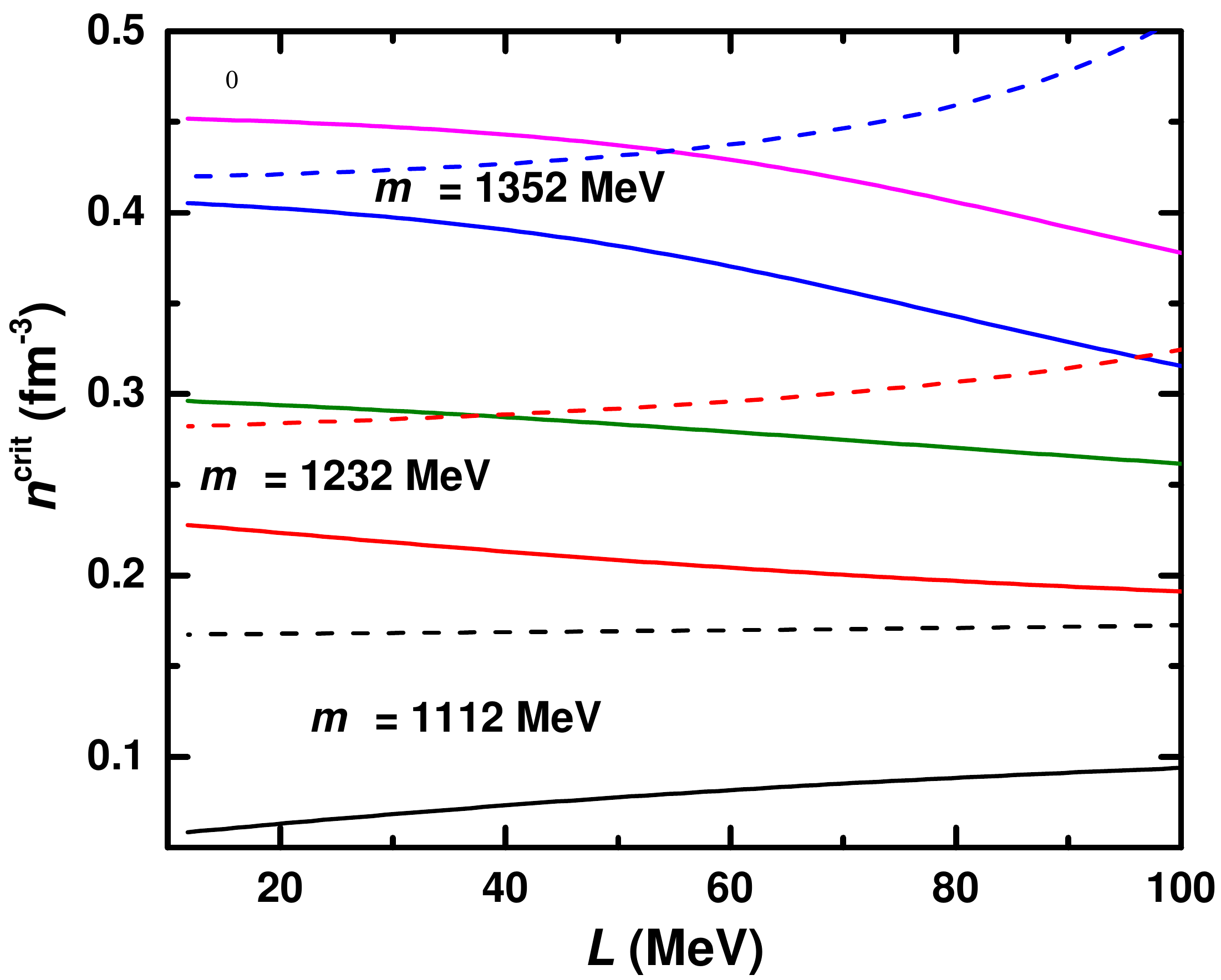}
\caption{\label{Fig:ncrit}(Color online)~The onset densities of hyperons and $\Delta$s in $\beta$-stable nuclear matter
as functions of the symmetry energy slope $L$. The black ($m_\Delta = 1112$ MeV), red ($m_\Delta = 1232$ MeV), and blue
($m_\Delta = 1352$ MeV) curves correspond to the onset densities of $\Delta^-$, where the solid ones and dashed ones are
obtained with $g_{\rho \Delta} = 0$ and $g_{\rho N}$, respectively.}
\end{figure}

At given total baryon number density $n$, the properties of neutron star matter can be obtained by fulfilling the conditions of baryon
number conservation with $n=\sum\limits_\mathrm{b}n_\mathrm{b}$, charge neutrality in Eq.~(\ref{eq:Chntr}), and chemical equilibrium
in Eq.~(\ref{eq:weakequi}) simultaneously. Similar to Ref.~\cite{Drago2014_PRC90-065809}, by varying the parameter $a_\rho$
in Eq.~(\ref{eq:ddcp_rho}), we examine the dependence of onset densities of $\Delta$s and hyperons $n_\mathrm{b}^\mathrm{crit}$ on the
symmetry energy slope $L$, which is fixed by fulfilling $\left.\mu_\mathrm{b}\right|_{\nu_\mathrm{b}=0}=\mu_n - q_\mathrm{b} \mu_e$.
A linear dependence of $L$ (in MeV) on $a_\rho$ is obtained, i.e., $L = 110.3 - 109.5 a_\rho$. The variation of $n_{\Lambda^0}^\mathrm{crit}$,
$n_{\Sigma^-}^\mathrm{crit}$, and $n_{\Delta^-}^\mathrm{crit}$ are presented in Fig.~\ref{Fig:ncrit}, while the onset densities for
other $\Delta$s and hyperons are much larger. For $\beta$-stable nuclear matter, the values of $\mu_e$ and $\rho_{0,3}$ are increasing with $L$.
Consequently, the obtained $n_{\Lambda^0}^\mathrm{crit}$ and $n_{\Sigma^-}^\mathrm{crit}$ are decreasing with $L$ while $n_{\Delta^-}
^\mathrm{crit}$ is increasing, which are consistent with the trends in \cite[Fig.~1]{Drago2014_PRC90-065809}. If we take $g_{\rho \Delta} = 0$,
the obtained $n_{\Delta^-}^\mathrm{crit}$ for $m_\Delta = 1232$ MeV and 1352 MeV are decreasing with $L$ since the contribution of
$\rho_{0,3}$ becomes irrelevant. Meanwhile, for the cases with $m_\Delta = 1112$ MeV, $n_{\Delta^-}^\mathrm{crit}$ is even smaller than
the saturation density. Since $\mu_e$ is decreasing with $L$ at subsaturation densities, the corresponding $n_{\Delta^-}^\mathrm{crit}$
(black solid curve) increases with $L$. Finally, it is worth mentioning that the variation of $n_{\Delta^-}^\mathrm{crit}$ with
respect to $L$ is insignificant comparing with $m_\Delta$ due to its relatively larger uncertainty.

\begin{figure}
\includegraphics[width=\linewidth]{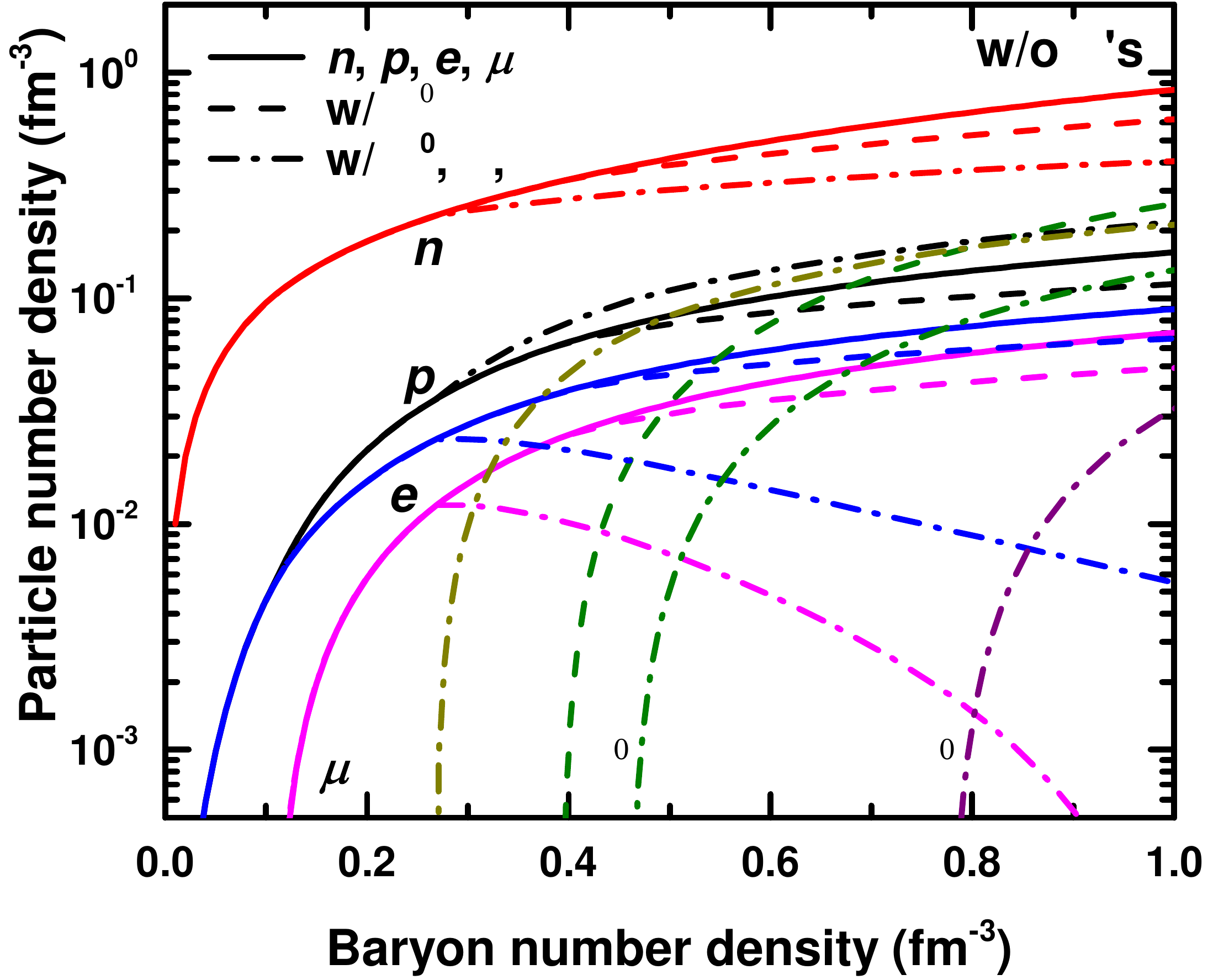}
\caption{\label{Fig:ni} (Color online)~Particle number densities for baryons and leptons in neutron star matter as functions of the total baryon
number density $n$ without $\Delta$ resonances.}
\end{figure}

\begin{figure*}[t!]
\includegraphics[width=0.8\linewidth]{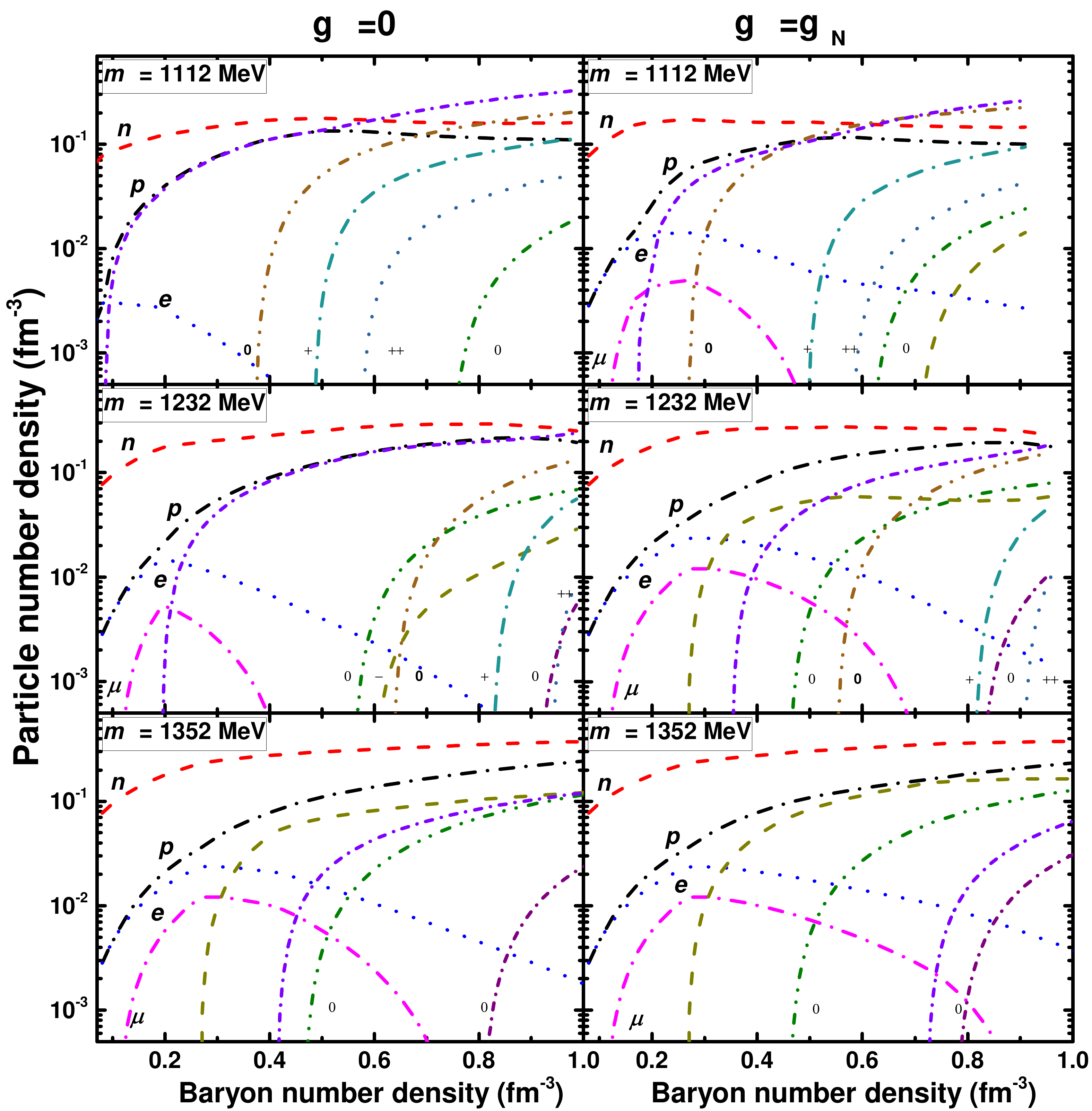}
\caption{\label{Fig:nid} (Color online)~Same as Fig.~\ref{Fig:ni} but including $\Delta$ resonances with $m_\Delta = 1112$~MeV, $1232$~MeV, and $1352$~MeV.}
\end{figure*}

The particle number density for each species is determined by Eq.~(\ref{eq:ni}), where the corresponding values are
presented as functions of the total baryon number density $n$ in Figs.~\ref{Fig:ni} and \ref{Fig:nid}. By including $\Lambda^0$
in nuclear matter, as indicated by the dashed curves in Fig.~\ref{Fig:ni}, the densities of protons and neutrons are slightly
reduced on the emergence of $\Lambda^0$. If we also include other hyperons such as $\Xi^{0,-}$ and $\Sigma^{+,0,-}$
(dash-dotted curves), since similar potential well depths are adopted for $\Lambda$'s and $\Sigma$'s, the $\Sigma^-$ firstly appears
at $n=0.27\ \mathrm{fm}^{-3}$ due to the negative charge it carries. In such cases, the number densities of leptons decrease
while those of protons increase. Meanwhile, the onset density of $\Lambda^0$ is increased from $n = 0.39\ \mathrm{fm}^{-3}$
to 0.46 $\mathrm{fm}^{-3}$ due to the inclusion of the negatively charged $\Sigma^-$. Since $\Xi^{0,-}$ possess the largest masses, their
onset densities are much larger with $n_{\Xi^-}^\mathrm{crit} = 1.2\ \mathrm{fm}^{-3}$ and $n_{\Xi^0}^\mathrm{crit}>n_{\Xi^-}^\mathrm{crit}$,
which exceed the density limit of Fig.~\ref{Fig:ni}.

The effects of $\Delta$ resonances are also studied and the results are shown in Fig.~\ref{Fig:nid}. To consider the
Breit-Wigner mass distribution of the $\Delta$ baryons and the possible in-medium mass shift~\cite{Cai2015_PRC92-015802},
three masses $m_\Delta = 1112$ MeV, 1232 MeV and 1352 MeV are adopted in our calculation. Note that the nucleon effective mass
$m_N^*\equiv m_N + g_{\sigma N}\sigma$ may become negative at higher densities. This is out of the scope of our current study
and we do not consider such cases. Thus, when we adopt $m_\Delta = 1112$ MeV, 1232 MeV and $g_{\rho \Delta} = g_{\rho N}$,
in Fig.~\ref{Fig:nid} we do not present the results with $m_N^*<0$ at the higher densities. For all $\Delta$ baryons, the negatively
charged $\Delta^-$ appears firstly as we increase the density. The onset density of $\Delta^-$ is found to increase both with $m_\Delta$
and $g_{\rho \Delta}$, which is consistent with previous findings~\cite{Cai2015_PRC92-015802, Zhu2016_PRC94-045803}. For massive
$\Delta$s ($m_\Delta = 1352$ MeV), the effects of $\Delta$ resonance are insignificant and only $\Delta^-$ appears. In the comparison with
hyperons, the massive $\Delta^-$ appears at larger densities than $\Sigma^-$, where the densities of hyperons are similar as the
cases in Fig.~\ref{Fig:ni}. If we adopt smaller values of $m_\Delta$ and $g_{\rho \Delta}$, the effects of $\Delta$ resonances
become important, where $\Delta^-$, $\Delta^0$, $\Delta^+$, and $\Delta^{++}$ appear sequentially as increasing the density.
Consequently, hyperons are hindered and appear only at larger densities. In the extreme case of $m_\Delta =1112$ MeV and
$g_{\rho \Delta} = 0$, the only left hyperon is $\Lambda^0$, which appears at a much larger density $n_{\Lambda^0}^\mathrm{crit}
= 0.74\ \mathrm{fm}^{-3}$. Note that a first-order phase transition from nuclear matter to $\Delta$ matter takes place in
the density range $n = 0.083$~-~$0.17\ \mathrm{fm}^{-3}$, where we have shown the corresponding densities in the lower left
panel of Fig.~\ref{Fig:nid}.

\begin{figure}
\includegraphics[width=\linewidth]{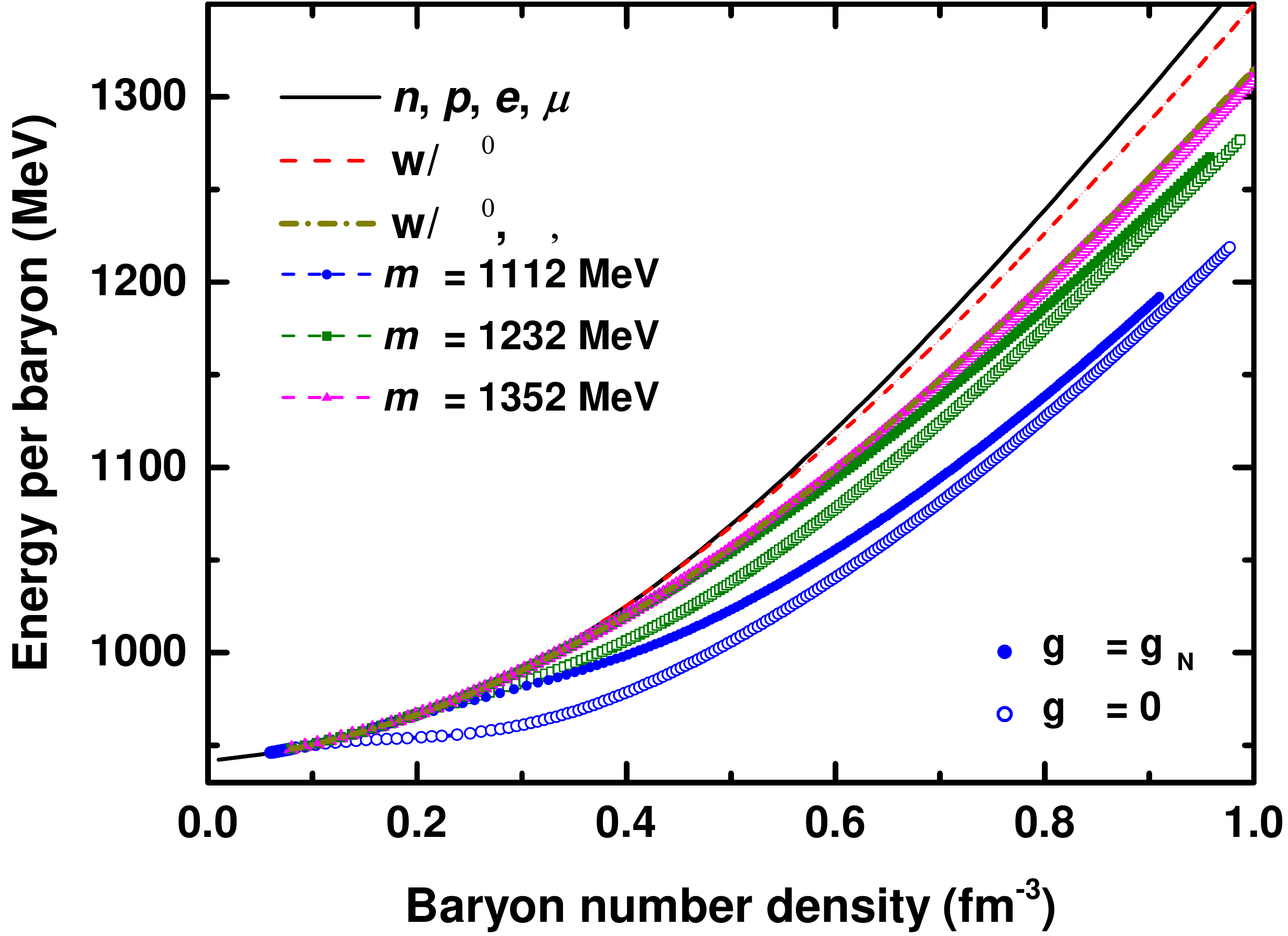}
\caption{\label{Fig:EpA}(Color online)~The energy per baryon of neutron star matter as functions of the baryon number density $n$.
The solid and open symbols are results obtained with $g_{\rho\Delta}=g_{\rho N}$ and 0, respectively. The same convention
is adopted for the following figures.}
\end{figure}

Based on the number density of each species, the energy density $E$ and pressure $P$ of neutron star matter can be obtained from
Eqs.~(\ref{eq:E}) and (\ref{eq:pressure}). In Fig.~\ref{Fig:EpA} we present the energy per baryon of neutron star matter as a
function of the baryon number density. As expected, the EoS becomes soft once we include new degrees of freedom. For hyperonic
matter (dash-dotted curve), if we consider $\Delta$ resonances and adopt the largest mass, i.e., $m_\Delta = 1352$ MeV, the EoS
is modified slightly at high density regions since only $\Delta^-$ appears at insignificant densities $n_{\Delta^-}$. Moreover,
adopting smaller values of $m_\Delta$ and $g_{\rho \Delta}$ would result in softer EoSs, where in the extreme case of $m_\Delta =1112$ MeV
and $g_{\rho \Delta} = 0$, a softest EoS is obtained for neutron star matter.

\begin{figure}
\includegraphics[width=\linewidth]{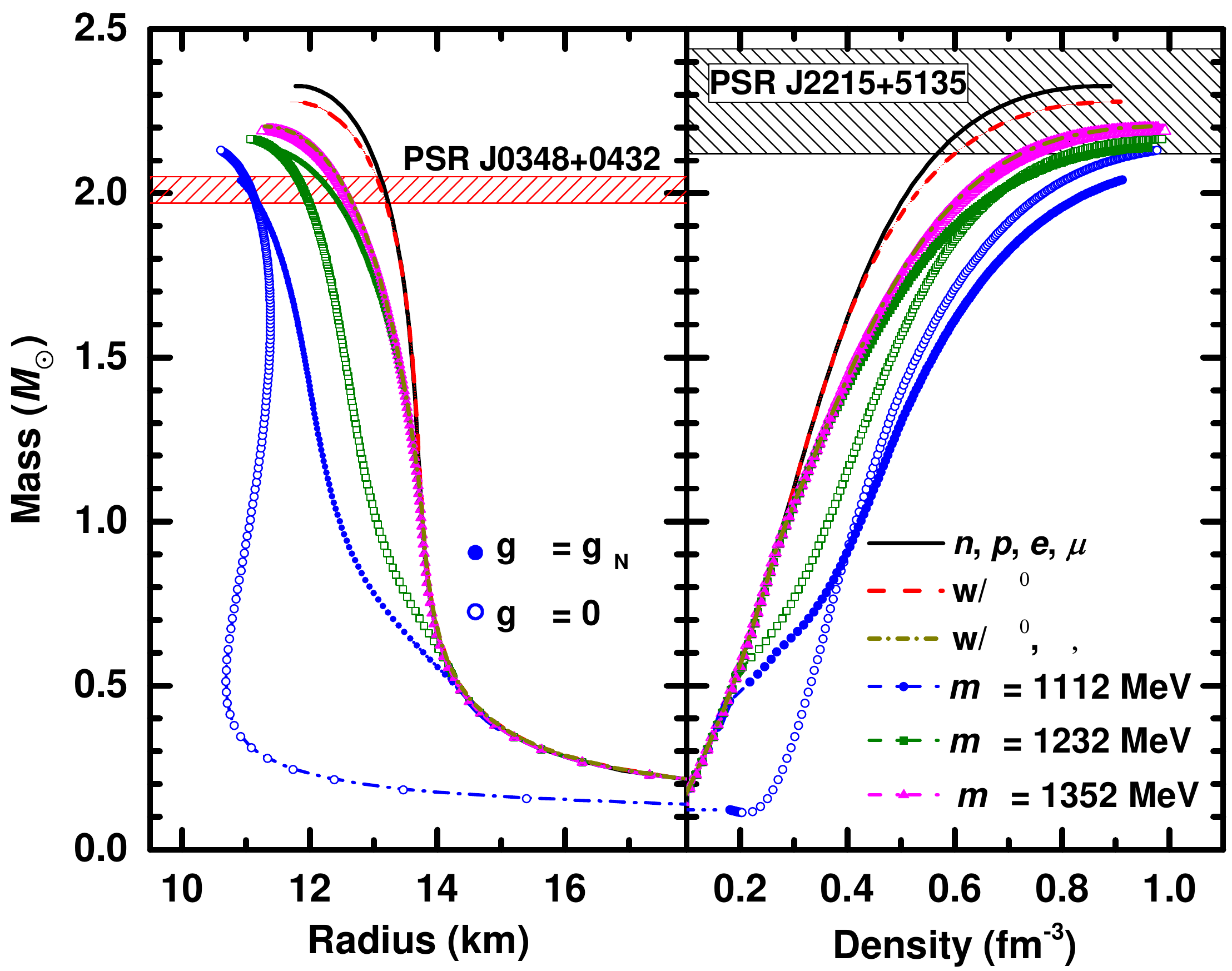}
\caption{\label{Fig:MR}(Color online)~The obtained mass-radius relations of compact stars including the possible existence of hyperons and $\Delta$
resonances. The masses of pulsars PSR J0348+0432 ($2.01 \pm 0.04\ M_\odot$)~\cite{Antoniadis2013_Science340-1233232} and PSR J2215+5135
($2.27{}_{-0.15}^{+0.17}\ M_\odot$)~\cite{Linares2018_ApJ859-54} are indicated with horizonal bands.}
\end{figure}

Based on the EoSs displayed in Fig.~\ref{Fig:EpA}, the structure of a neutron star can be determined by solving the TOV equation
in Eq.~(\ref{eq:TOV}). For neutron star matter at subsaturation densities ($n\leq 0.08\ \mathrm{fm}^{-3}$), we adopt the EoS
presented in Refs.~\cite{Feynman1949_PR75-1561, Baym1971_ApJ170-299, Negele1973_NPA207-298}, where the properties of crystalized
matter that forms the neutron star crust can be well described. In Fig.~\ref{Fig:MR} we show the masses of compact stars as
functions of radius (Left panel) and central baryon number density (Right panel), where the possible existence of hyperons and $\Delta$
resonances are considered. The obtained results are compared with the observational masses of PSR J0348+0432 ($2.01 \pm 0.04\
M_\odot$)~\cite{Antoniadis2013_Science340-1233232} and PSR J2215+5135 ($2.27{}_{-0.15}^{+0.17}\ M_\odot$)~\cite{Linares2018_ApJ859-54}.
As we include more degrees of freedom, the maximum mass and radii of compact stars become smaller. For compact stars
including $\Delta$ resonances, if we adopt $m_\Delta =1112$ MeV and $g_{\rho \Delta} = g_{\rho N}$, the maximum mass does not reach
the lower limit of PSR J2215+5135. This can be fixed by using smaller values of $\rho$-$\Delta$ couplings, e.g., $g_{\rho \Delta}
= 0$. Due to the occurrence of a first-order phase transition at small densities ($n = 0.083$-$0.17\ \mathrm{fm}^{-3}$),
a smallest radius with $R=11.3$ km for 1.4 $M_\odot$ compact star is obtained, which is consistent with the recent measurements
of neutron star radii~\cite{Guillot2013_ApJ772-7, Lattimer2014_EPJA50-40, Ozel2016_ARAA54-401, Li2015_ApJ798-56,
Steiner2018_MNRAS476-421, LigoVirgo2018_arXiv1805.11581}.

\begin{figure}
\includegraphics[width=\linewidth]{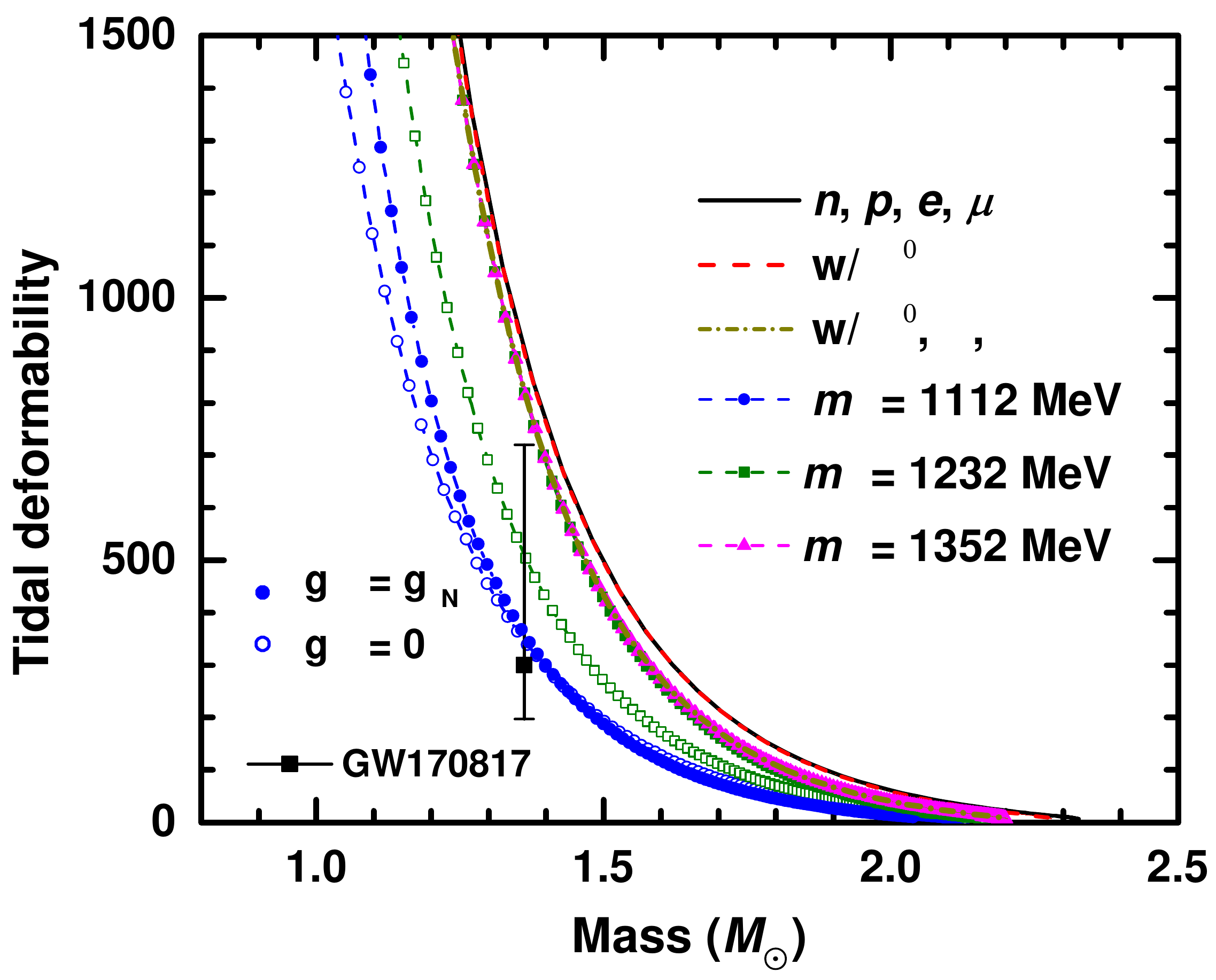}
\caption{\label{Fig:Lambda}(Color online)~The tidal deformabilities of compact stars as functions of their masses. The recent constraint obtained
with the binary neutron star merger event GW170817 is indicated with the black solid box~\cite{LIGO2017_PRL119-161101, LigoVirgo2018_arXiv1805.11579,
Coughlin2018_arXiv1805.09371}.}
\end{figure}

Another important constraint is the tidal deformability of the compact stars,  which can be obtained based on
Eq.~(\ref{eq:td}). In Fig.~\ref{Fig:Lambda} we present the tidal deformabilities of compact stars corresponding to those
in Fig.~\ref{Fig:MR}. The observation of binary neutron star merger event GW170817 have set the dimensionless combined tidal
deformability $197 \leq \bar{\Lambda}\leq 720$~\cite{LigoVirgo2018_arXiv1805.11579, Coughlin2018_arXiv1805.09371}, which
is a mass-weighted linear combination of tidal deformabilities~\cite{Favata2014_PRL112-101101}
\begin{equation}
  \bar{\Lambda} = \frac{16}{13} \frac{(m_1+12 m_2)m_1^4\Lambda_1+(m_2+12 m_1)m_2^4\Lambda_2}{(m_1+m_2)^5}.
\end{equation}
Since $\bar{\Lambda}$ is insensitive to the mass ratio $m_2/m_1$~\cite{Bhat2018_arXiv1807.06437}, combined with the best measured
chirp mass $\mathcal{M} = {(m_1 m_2)^{3/5}}{(m_1+m_2)^{-1/5}}=1.186\pm 0.001 M_\odot$~\cite{LigoVirgo2018_arXiv1805.11579},
in Fig.~\ref{Fig:Lambda} we show the corresponding constraint on the tidal deformability $\Lambda=\Lambda_1=\Lambda_2$ at
$m_1=m_2=1.362 M_\odot$. It is found that the observational tidal deformability has put a strong constraint on the compositions
of compact stars, so that the $\Delta$ resonances have to be included. Meanwhile, as discussed before, a small enough
$\Delta$-$\rho$ coupling $g_{\rho \Delta}$ should also be adopted for compact stars to reach the mass of PSR J2215+5135.

\section{\label{sec:con}Conclusion}
We explore the possible existence of hyperons and $\Delta$ resonances in compact stars. The properties of baryonic
matter is obtained based on the RMF models. For the $N$-$N$ interactions, we adopt the covariant density functional
PKDD~\cite{Long2004_PRC69-034319}, while the hyperon-meson couplings are fixed based on our previous investigations on
hypernuclei and neutron stars~\cite{Sun2018_CPC42-25101, Liu2018_PRC98-024316}. For the $\Delta$-meson couplings, we adopt the
universal baryon-meson coupling scheme. Meanwhile, to consider the possibility of smaller $g_{\rho \Delta}$ and mass
variations, we also study the cases with $g_{\rho \Delta}=0$ and various $\Delta$ masses with $m_\Delta =1112$~MeV, $1232$~MeV, and $1352$~MeV.
The EoSs of neutron star matter become softer once we include new degrees of freedom. By solving the TOV equation with
these EoSs, we obtained the masses, radii, and tidal deformabilities of the corresponding compact stars. Comparing with
the dimensionless combined tidal deformability $197 \leq \bar{\Lambda}\leq 720$ constrained according to the recent
observations of GW170817~\cite{LigoVirgo2018_arXiv1805.11579, Coughlin2018_arXiv1805.09371}, we find it is essential
to include the $\Delta$ resonances in compact stars, and the $\Delta$-$\rho$ coupling $g_{\rho \Delta}$ should be small
enough if the mass of PSR J2215+5135 ($2.27{}_{-0.15}^{+0.17}\ M_\odot$)~\cite{Linares2018_ApJ859-54} is confirmed.

\section*{ACKNOWLEDGMENTS}
This work was supported by National Natural Science Foundation of China (Grant Nos.~11375022, 11475110, 11525524, 11505157, 11575190,
11705163, 11775014, 11711540016, and 11621131001), and the Physics Research and Development Program of Zhengzhou University (Grant No.~32410017).
The computation for this work was supported by the HPC Cluster of SKLTP/ITP-CAS and the Supercomputing Center, CNIC, of the CAS.

\newpage
%%
%% reference here
%%
%\bibliography{strange_quark}

%

\end{document}